\documentstyle[preprint,aps,eqsecnum]{revtex}
 
\input{psfig}
\def\be{\begin{equation}}
\def\ee{\end{equation}}
\def\bea{\begin{eqnarray}} 
\def\eea{\end{eqnarray}}
\def\line{\hbox to \hsize}    
\def\frac #1#2{{#1\over #2}}

\def\bpsi{{\overline \psi}}
\def\bzeta{{\overline\zeta}}

\def\bxi{{\overline\xi}}
\def \z{{\overline z}} 
\def \bz{{\overline z}} 

\def\ad{ a^{\dagger}} 
\def\bd{ b^{\dagger}}    

\def\hH{{\hat H}}

\def\Det{{\rm Det\,}}
 
\def \B{{\overline B}}

\def \ket #1{{\vert #1\rangle}}
\def \bra #1{{\langle #1\vert}}
\def \brak #1#2{{\langle#1\vert#2\rangle}}
\def\eval #1#2#3{{\langle#1\vert#2\vert#3\rangle}} 

\def\1{\mbox{\bf 1}}
\def\nnu{\nonumber}
\def\lf{\left}
\def\rt{\right}
\def\line{\hbox to \hsize}    
\def\eno#1{Eq.~(\ref{#1})}

\def\eps{\epsilon}
\def\sech{{\rm sech}}
\def\jtil{{\tilde\jmath}}
\def\htil{{\tilde h}}
\def\ktil{{\tilde k}_1}
\def\Rtil{{\tilde R}}
\def\Lam{\Lambda} 
\def\lam{\lambda}

\def\hf{{1\over 2}}
\def\tshf{{\textstyle{1\over 2}}}
\def\zcl{z_{\rm cl}}
\def\bzcl{\bz_{\rm cl}}
\def\rtl{\sqrt\lam}
\def\rtlb{\sqrt{1-\lam}}
\def\rtb{\sqrt{1-\htil^2}}
\def\rthb{\sqrt{1-h^2}}
\def\root2{\sqrt{2}}
\def\rat#1{\lf({1 - v_{#1} \over -1 - v_{#1}} \rt)}
\def\Fe8{{Fe$_8$}}
\def\yhat{{\hat y}}
\def\mel#1#2#3{{\langle#1\vert#2\vert#3\rangle}}
\begin{document}
\draft %(only for revtex) 

\title{\bf SPIN  COHERENT-STATE PATH
INTEGRALS   AND
THE INSTANTON CALCULUS}

\author{ANUPAM GARG}
\address{ Northwestern
University, Department of Physics and Astronomy\\Evanston,
IL 60208 USA
%\\E-mail agarg@nwu.edu
}

\author{EVGUENY KOCHETOV}
\address{Bogoliubov Theoretical Laboratory,\\
Joint Institute for Nuclear Research, 141980
Dubna, Russia} 

\author{KEE-SU PARK}
\address{Department of Physics,\\ 
Pohang University of Science and Technology\\
Pohang, Kyungbuk, Korea}

%\address{University of Illinois, Department of Physics\\ 1110 W. Green St.\\
%Urbana, IL 61801 USA
%\\E-mail: k-park4@uiuc.edu
%} 

\author{MICHAEL STONE}
\address{University of Illinois, Department of Physics\\ 1110 W. Green St.\\
Urbana, IL 61801 USA
%\\E-mail: m-stone5@uiuc.edu
}

\maketitle

\vfil\eject

\begin{abstract}

We use an instanton approximation  to the   continuous-time
spin coherent-state path integral to obtain  the  tunnel
splitting of classically degenerate ground states. We show
that provided  the fluctuation determinant is carefully 
evaluated, the path integral expression is accurate to
order $O(1/j)$. We apply the method to the LMG model and to
the molecular magnet  Fe$_8$ in a transverse field. 

\vskip 72pt
\noindent {\it This paper is dedicated to the memory of\/} {\bf Victor
Belinicher,\/}  {\it who was lost 
when Siberia Airlines flight 1812 was shot
down  over the Black Sea, Oct
4th 2001. Victor made many contributions to physics, in particular to
the spin tunneling problem.\/}

\end{abstract}
\vfill

%\newpage 

\section{Introduction}

One of the most convincing demonstrations of quantum
effects  in a near-macroscopic system is provided by the
accurate measurement\cite{wernsdorfer99} of the level splittings in the
molecular magnet Fe$_8$. These splittings are caused by the
quantum tunnelling of the  direction of the large ($J=10$) 
molecular spin between two classically degenerate energy minima. 

The  natural tool for studying such  spin tunnelling should
be the   spin ($SU(2)$)  coherent-state path integral.  It
is easy to establish that this  formalism  gives a good 
qualitative description of the tunnelling process
\cite{chudnovsky88,loss92,henley92} --- including the
dramatic  topological quenching of the  tunnelling
\cite{garg93} that makes the  $Fe_8$ results so
interesting. Unfortunately,   a straightforward application
of the spin coherent-state path integral to compute the
semiclassical propagator \cite{kuratsuji81} or the tunnel
splitting \cite{garg_kim92} yields results that are
incorrect beyond the leading exponential order. 

Although there do exist other path integral approaches
which find the splitting correctly
\cite{enz86,belinicher97}, the resulting calculations tend
to be  intricate, and the simplicity seen in the
conventional Schr{\"o}dinger particle case is lost. These
problems have lead to the spin coherent state path integral
acquiring a reputation for being mathematically ill
defined  --- or at least harder to deal with than the
conventional Feynman path integral, whose mathematical
subtleties have been well studied. 

Recently, however, it has begun to be appreciated that the
problem with the spin coherent state calculation is simply
that the fluctuation determinant has an ``anomaly'', and
that, once the ``extra phase''  provided by the anomaly  
is taken into account, the coherent state  path integral
gives correct answers.  This extra phase seems to have
been originally   discovered in the 1980's by Solari
\cite{solari87}, but the significance of his result was not
widely appreciated. It was then rediscovered by  one of 
the present authors  \cite{kochetov95} and also by
Vieira and   Sacramento\cite{sacramento95}. The
interpretation of the extra phase as an ``anomaly'' is due
to the remaining authors of the present
paper\cite{stone00}.

These previous discussions of the extra phase were
restricted to the case of quantum evolution between generic
values of the classical degrees of freedom. However, when
we calculate tunnel splitting, the endpoints of the 
instanton path lie at  local minima of the classical
energy  and, just as in the Schr{\"o}dinger particle case,
the Jacobi fluctuation operator has a zero mode which 
makes the inverse of its  determinant singular and the
general formula for the propagator inapplicable. Thus our
earlier work had not fully established that the spin coherent state path
integral gave the correct result for the tunnel splitting.
This we do in the present paper.

In the next section we provide a brief review of the spin
coherent-state path integral, including the correction to
the  fluctuation determinant prefactor.
In sections three and  four we discuss the complications that ensue when
there is a zero mode and provide a general formula for the
one-instanton contribution to the tunnelling amplitude.
In section  five  we apply this formula to the relatively simple case of
the Lipkin-Meshkov-Glick (LMG) model\cite{lipkin65}, and in
section six we
evaluate the tunnel splitting for a realistic model of Fe$_8$.

\section{Spin Coherent States}

We follow the conventions in \cite{stone00} and define our spin coherent
states\cite{perelomov86} to be  
\be
\ket{z} = \exp(z \hat J_+) \ket{j,-j},
\ee
where $\ket{j,-j}$ is the lowest  spin state in the $2j+1$
dimensional representation of $SU(2)$ and 
$\hat J_+$ is the spin algebra  ladder operator
obeying
\be
\hat J_+ \ket{j,m} = \sqrt{j(j+1)-m(m+1)}\ket{j,m+1}.
\ee
The variable $z$
is  a stereographic coordinate on the
unit sphere with $z=0$ at  the south pole (spin down
direction) and
$z=\infty$ at the north pole (spin up). 

These coherent states are  not normalized, but depend  
holomorphicly on $z$. This means that  matrix elements 
such as $\eval{z'}{\hat O}{z}$ are  holomorphic functions of the
variable $z$, and anti-holomorphic functions of the
variable $z'$.

The inner product of two coherent  states is  
\be
\brak{ z'}{z} = (1+ \z' z)^{2j},
\ee
and they satisfy the   overcompleteness relation    
\be
\1 = \frac {2j+1}{\pi} \int \frac{d^2z}{(1+\z z)^{2j+2}}
\,{\ket{z}} {\bra{z}}.\
\label{EQ:overcompleteness}
\ee
Here  $d^2z$ is shorthand for  $dx\,dy$. The
factor  $1/(1+\z z)^2$ combines with this to make the
invariant measure on the the two-sphere.  
The remaining factor in the integration measure,
$1/(1+\z z)^{2j}$, serves to normalize the  coherent states.

%The overcompleteness provides the inner product for the
%coherent-state 
%wavefunctions.
%If $f(z) =\brak{f}{z}$ and $g(z)=\brak{g}{z}$ then
%\bea
%\brak{f}{g} &=&  
%\frac {2j+1}{\pi} \int \frac{d^2z}{(1+\z z)^{2j+2}}\,\brak{f}{z}\brak{z}{g}
%\nonumber\\&=& 
%\frac {2j+1}{\pi} \int \frac{d^2z}{(1+\z z)^{2j+2}}\,\overline{
%g(z)}f(z).
%\eea

We may use the overcompleteness relation to derive a 
formal continuous-time  path integral representation for the propagator 
\be
K(\bzeta_f,\zeta_i,T)= \eval{\zeta_f}{e^{-i\hat H
T}}{\zeta_i}.
\label{EQ:propagator}
\ee
We insert   $N$ intermediate overcompleteness relations into
(\ref{EQ:propagator}) and consider the 
limit $N\to \infty$.  This leads to the path integration
formula\cite{kochetov95}
\be
K(\bzeta_f,\zeta_i,T)=\int_{\zeta_i}^{\bzeta_f}d\mu(\z,z)
\exp\{S(\z(t), z(t))\},
\ee
where the path measure $d\mu$ is 
\be
d\mu(\z(t),z(t))=\lim_{N\to\infty} \prod_{n=1}^N\frac{2j+1}{\pi}
\frac{d^2z_n}{(1+\z_n z_n)^{2}},
\ee
and the 
action $S(\z(t),z(t))$ is   
\be
S(\z(t),z(t)) = j\left\{\ln(1+ \bzeta_f z(T)) 
+\ln(1+ \z(0)\zeta_i)\right\}
+
\int_0^T \left\{ j \frac{\dot\z z- \z \dot z}{1+\z z} 
-i H(\z,z)\right\} dt.
\ee
The c-number Hamiltonian, $H(\z,z)$, is obtained from  the
operator $\hH$ by 
\be
H(\z,z)= \eval{z}{\hH}{z}/\brak{z}{z}.
\ee
The paths $z(t)$, $\z(t)$ obey the boundary conditions
$z(0)=\zeta_i$, $\z(T)=\bzeta_f$, but $\z(0)$, $z(T)$, being
actually $\z(0+\epsilon)$ and $z(T-\epsilon) $, are unconstrained, 
and are to be integrated over\cite{kochetov95}.  

The manipulations leading to the continuous time path
integral are heuristic,  but with careful treatment the
formal path integral should be as useful as the familiar
configuration space Feynman path integral.   In particular
the semiclassical, or large $j$,  propagator can be 
obtained from a stationary phase approximation to the path
integral \cite{stone00}. 

The stationary phase approximation   requires us to  seek
``classical'' trajectories  for which  $S$ remains
stationary as we vary the   functions $z(t)$ and $\z(t)$. 
These stationary paths will generally be  complex.  If we write 
$z$ as $x+iy$ and $\z=x-iy$, then, except in special cases,
$x$ and $y$ are not real numbers. In
particular there is no requirement that $\z(0)$  be  the complex
conjugate of $z(0)\equiv\zeta_i$, nor that   $z(T)$ be the
complex conjugate of $\z(T)\equiv \bzeta_f$.
Bearing this in mind, we make variations about a chosen path, 
and  keep track of all boundary contributions
resulting from integrations by parts. We find that
\bea 
\delta S &=&
\frac{2j z(T)}{1+ \bzeta_f z(T)}\delta \z(T) + 
\frac{2j \z(0)}{1+ \z(0)\zeta_i} \delta z(0)\nonumber\\
&&+ 
\int_0^T \left\{ 
\delta z(t)\left( \frac {2j\dot\z}{(1+\z z)^2}-
i \frac{\partial H}{\partial z}\right)
+ 
\delta \z(t)\left( -\frac {2j\dot z}{(1+\z z)^2}-
i \frac{\partial H}{\partial \z}\right)\right\}dt.
\label{EQ:hamilton-jacobi}
\eea
Demanding that  this change in the action be zero
requires  the  trajectory  to obey
Hamilton's equations
\be
\dot\z =i \frac {(1+\z z)^2}{2j}\frac{\partial H}{\partial z},
\quad 
\dot z =- i \frac {(1+\z z)^2}{2j}\frac{\partial H}{\partial \z},
\label{EQ:hamilton}
\ee
together with the conditions  $\delta z(0)=0$ and $\delta
\z(T)=0$. We can therefore  impose   the boundary
conditions   $z(0)=\zeta_i$,  $\z(T)=\bzeta_f$, but 
$\z(0)$ and $z(T)$  are free to  vary, and so are
determined by the equations of motion. This is
important because Hamilton's equations are first order in
time and we cannot simultaneously impose initial and final
conditions on their solutions. 

The dynamically determined endpoints can  also be read off from  the
Hamilton-Jacobi relations  that follow from  
(\ref{EQ:hamilton-jacobi}). These are   
\be
\frac{\partial S_{\rm cl}}{\partial \bzeta_f}= \frac {2jz(T)}{1+ \bzeta_f z(T)},
\quad
\frac{\partial S_{\rm cl}}{\partial \zeta_i}= \frac {2j\z(0)}{1+
\z(0)\zeta_i}.
\label{EQ:jacobi_equations}
\ee

The Hamilton-Jacobi relations also tell us that  
\be 
\frac{\partial S_{\rm cl}}{\partial \bzeta_i}=
\frac{\partial S_{\rm cl}}{\partial
\zeta_f}=0,
\ee
showing that $S_{\rm cl}$ is a holomorphic function of
$\zeta_i$, and an anti-holomorphic function of $\zeta_f$.
These  analyticity properties of $S_{\rm cl}$ coincide with
those of $K$. This is reasonable since $\exp S_{\rm cl}$ 
is the leading approximation to $K$, and we would expect
analyticity to be preserved term-by-term in the large $j$
expansion. Finally, we have the Hamilton-Jacobi equation 
\be  
\frac{\partial S_{\rm cl}}{\partial T}= -i H(\bzeta_f,
z(T)).
\ee

In \cite{stone00} we showed that after we compute
the Gaussian integral over small fluctuations about the
stationary phase path the resulting
semiclassical approximation to the propagator is  
\be
K_{\rm scl} (\bzeta_f,\zeta_i,T)=
\left(\frac{(1+\bzeta_f z(T))(1+\z(0)\zeta_i)}{2j}
\frac {\partial^2 S_{\rm cl}}{\partial \zeta_i\partial
\bzeta_f}\right)^{\frac 12} \exp\left\{S_{\rm cl}(\bzeta_f, \zeta_i, T)
+\frac{i}{2}\int_0^T \phi_{\rm SK}(t)dt\right\},
\label{EQ:kochetov_propagator}
\ee
or a sum of such terms over a set of contributing classical
paths. In this expression   
\be 
\phi_{\rm SK}(\z,z)=  \frac 12 \left(\frac{\partial}{\partial \z} \frac{(1+ \z z)^2}{2j}
\frac{\partial H}{\partial z} + \frac{\partial}{\partial z} \frac{(1+ \z z)^2}{2j}
\frac{\partial H}{\partial \z}\right),
\ee
is the ``extra-phase'' discovered by Solari, Kochetov, and
Vieira and Sacramento.

The form (\ref{EQ:kochetov_propagator}) is valid only
if the prefactor is finite. When we compute instanton
contributions to tunnelling there is a zero mode in the
quadratic form for  small fluctuations, and the resulting divergent
integral over this mode is to be replaced by an integral
over a  collective coordinate labeling  the  
instant  that the tunnelling event occurred. This we will describe
in the next section.

\section{Dealing with the Zero Mode}

As is usual in calculating tunnelling effects, it is convenient to
perform the computations in Euclidean (imaginary) time. For
the sake of symmetry we will take the time evolution as running from  
$-T/2$ to $T/2$
and  the  propagator (\ref{EQ:kochetov_propagator})
becomes 
\be
K(\bzeta_f, \zeta_i, T) = \left[ D(T)\right]^{-\tshf}\exp\left\{S_{\rm
cl} + \tshf \int_{-T/2}^{T/2} \phi_{\rm
SK} d\tau\right\}, 
\ee where  again $\phi_{\rm SK}$ is the integrand of the Solari-Kochetov phase
\be
\phi_{\rm SK} = \frac 12 \left(\frac{\partial}{\partial \z} \frac{(1+ \z z)^2}{2j}
\frac{\partial H}{\partial z} + \frac{\partial}{\partial z} \frac{(1+ \z z)^2}{2j}
\frac{\partial H}{\partial \z}\right), \label{EQ:defsk}
\ee
evaluated along $\zcl(\tau)$, $\bzcl(\tau)$, and  $D(T)$ is
the fluctuation determinant. The latter   may be  found by the
``shooting method''.  As explained in \cite{stone00}, this involves 
solving the equation 
\be
\hat L \Psi_L\equiv\left[\matrix{ B(\tau) & -\partial_\tau + A(\tau) \cr
              \partial_\tau + A(\tau) &
	      \B(\tau)}\right] 
\left(\matrix{ \psi_L \cr
               \bpsi_L}\right) =0,
\label{EQ:jacobi}
\ee	       	        
where 
\bea
 A =\phi_{\rm SK} &=& \frac 12 \left(\frac{\partial}{\partial \z} \frac{(1+ \z z)^2}{2j}
\frac{\partial H}{\partial z} + \frac{\partial}{\partial z} \frac{(1+ \z z)^2}{2j}
\frac{\partial H}{\partial \z}\right), \nonumber\\
B &=& \frac{\partial}{\partial \z} \frac{(1+ \z z)^2}{2j}
\frac{\partial H}{\partial \z}, \nonumber\\
\B &=& \frac{\partial}{\partial z} \frac{(1+ \z z)^2}{2j}
\frac{\partial H}{\partial z},
\eea
with the initial condition
\be
\Psi_L(-T/2)=\left(\matrix{ \psi_L \cr
               \bpsi_L}\right)_{-T/2} =\left(\matrix{ 0 \cr
               1}\right).
\ee 
Given  the solution of this equation, 
we read off  the determinant as $D(T) = \bpsi_L(T/2)$. In real
time, and when there are no problems with zero-modes, 
this recipe leads to the prefactor appearing in
(\ref{EQ:kochetov_propagator}).
 
Now assume that the coherent states  $\ket{z_i}$ and
$\ket{z_f}$ represent  spins pointing  along the directions
of  two equal-energy  global minima of the Hamiltonian
$\hH$. Because the  gradient of the energy vanishes at both
ends, the classical path joining $z_i$ to $z_f$ has the
character of an instanton: as  the total time taken to
traverse the path becomes longer and longer most of the
motion still takes place in an ``instant'', a fixed period
short in duration compared to the total. When $T$ becomes
infinite, the   epoch of this ``instant''  is
arbitrary and this leads to a
zero-eigenvalue mode in the fluctuation operator.  Thus
$D(T)$ is formally zero. The problem of dividing by the
square root of zero is avoided by introducing a collective
coordinate for the tunnelling epoch, and the formal
infinity in the one-instanton contribution to the
propagator becomes a factor of $T$. 

The classical instanton solution can be written 
$\zcl(\tau-\tau_0)$,
$\bzcl(\tau-\tau_0)$ where $\tau_0$ is the epoch at which
the tunnelling occurs. Since, in the large $T$ limit, the
action for the tunnelling event is independent of $\tau_0$,
the normalized zero mode is 
\be
\Psi_0 = \left(\matrix{ \psi_0(\tau)\cr
                        \bpsi_0(\tau)}\right) =
\frac{\sqrt{g}}{1+\bz_{\rm cl}z_{\rm cl}} 
\left(\matrix{ \dot \zcl(\tau)\cr 
              \dot\bzcl(\tau)}\right), 
\ee
where $g$ is chosen to make    
\be
\int_{-T/2}^{T/2}\Psi_0^t\Psi_0\, d\tau= \int_{-T/2}^{T/2}\left(\psi_0^2 +\bpsi_0^2\right)d\tau =1.
\ee

The divergent Gaussian integration over the coefficient of the 
zero mode is replaced by an integral over possible tunnelling
epochs $\tau_0$ by inserting a factor of 
\be
1= \frac 1{\sqrt{2\pi \alpha}} \int_{-T/2}^{T/2} d\tau_0 \left(\frac{\partial
{\cal F}}{\partial \tau_0}\right)  \exp - \frac
{1}{2\alpha}{\cal F}^2(\tau_0)
\ee
into the path integral, with the choice 
\be
{\cal F}(\tau_0) = \int_{-T/2}^{T/2}
d\tau'\frac{1}{1+\bz_{\rm cl}z_{\rm cl}(\tau'-\tau_0)}
\Psi_0^t(\tau'-\tau_0) 
\left(\matrix{  z(\tau')\cr 
               \z(\tau')}\right),
\ee
and then proceeding in a manner similar to that used for
quantum mechanical instantons in the Feynman path
integral\cite{lowe78,collins78}: we first set $z= z_{\rm
cl}(\tau-\tau_0) + \delta z(\tau-\tau_0)$ and similarly
$\z$.
Next, after observing
that everything depends only on the combination
$\tau-\tau_0$,
we  change variables $\tau-\tau_0\to \tau$.  The integral 
over $\tau_0$ is then trivial and gives
a factor of $T$. Meanwhile, after an integration by parts
and ignoring the fuctuations of $(z, \z)$ about $(z_{\rm
cl}, \z_{\rm cl})$ which are of higher order,   
the Jacobian factor becomes   
\be
\frac{\partial
{\cal F}}{\partial \tau_0} = \int_{-T/2}^{T/2} d\tau'
\Psi_0^t\frac{1}{1+\bz_{\rm cl}z_{\rm cl}}\left(\matrix{ 
\dot z_{\rm cl}(\tau')\cr 
               \dot \z_{\rm cl}(\tau')}\right) = \frac 1{\sqrt g}.
\ee	       
The quadratic term in the exponent is a projector onto the
zero mode and  replaces the
vanishing  eigenvalue by $1/2j\alpha$. The net result  
is the replacement
\be
\left[D(T)\right]^{-\tshf} \to T\sqrt{\frac{j}{\pi g}} 
\left[\frac{D(T)}{\lambda_0}\right]^{-\tshf},
\label{EQ:collective}
\ee
where $\lambda_0(T)$ is the eigenvalue that vanishes as 
$T$ becomes large.

The desired ratio,
$\Det'(\hat L)={D(T)}/{\lambda_0}$, is equal to  
$\bpsi_L(T/2)/\lambda_0(T)$. We will not have to  
obtain $\bpsi_L(T/2)$ and $ \lambda_0(T)$
separately. 

The eigenvalue problem is 
\be
\hat L \Psi_{\lambda}=\lambda \Psi_\lambda; \quad
\Psi_\lambda= \left(\matrix{ \psi_\lambda \cr
\bpsi_\lambda}\right),
\label{EQ:eigenvalue_eq}
\ee where $\hat L$ is the same operator as in 
(\ref{EQ:jacobi}), but 
with boundary conditions
$\psi_\lambda(-T/2)=\bpsi_\lambda(T/2)=0$. 

For finite $T$ the shooting method solution, $\Psi_L$, is
close to, but  not quite equal to, the ``small-eigenvalue''
eigenfunction, $\Psi_{\lambda_0}$. Although $\Psi_L$ obeys
the boundary condition at $\tau=-T/2$, it does not quite obey
the boundary condition at $\tau =+T/2$. In turn 
$\Psi_{\lambda_0}$ is close to, but not quite equal to, the
infinite$-T$  zero-eigenvalue mode, $\Psi_{0}$. 

Now  $\Psi_{0}$ obeys the equation $\hat L\Psi_{0}=0$, but
no particular boundary conditions at $\pm T/2$. There is a 
second solution of this equation, $\Xi_0 = (\xi_0,
\bxi_0)^t$. The Wronskian of these solutions  
\be
W(\Psi,\Xi)= \left|\matrix {\psi_0(\tau)& \xi_0(\tau)\cr
                           \bpsi_0(\tau)&\bxi_0(\tau)}\right| 
\ee 
is independent of $\tau$. Next we observe that the differential
equation (\ref{EQ:eigenvalue_eq}) can be converted to an
integral equation
\bea
\Psi_\lambda(\tau) &=& \Psi_L(\tau) + {\lambda}
\int_{-T/2}^{\tau}
d\tau'G(\tau,\tau')\Psi_\lambda(\tau'),\nonumber\\ 
&=& \Psi_L(\tau) + \frac {\lambda}{W}
\int_{-T/2}^{\tau} d\tau' \left[ \Psi_0(\tau)\Xi_0^t(\tau')-
\Xi_0(\tau)\Psi_0^t(\tau')\right] \Psi_\lambda(\tau').
\label{EQ:integral_equation}
\eea  
Since $\Psi_L(\tau)$ obeys the boundary conditions at
$-T/2$, and the integral vanishes at this point,
we can find the
eigenvalues  $\lambda$ by requiring that the  lower component
at of $\Psi_{\lambda}$ vanishes at $\tau = T/2$.
We  are only interested in solutions where
$\lambda=\lambda_0$ is
very small. Because of this we can approximate the   
$\Psi_\lambda(\tau')$ appearing in the integral in
(\ref{EQ:integral_equation}) by  the
zeroth-order solution, $\Psi_L$. In this way we see that 
\be
\frac{\bpsi_L(T/2)}{\lambda_0(T)} = - \frac{1}{W}
 \int_{-T/2}^{T/2} d\tau \left[ \bpsi_0(T/2)\Xi_0^t(\tau)-
\bxi_0(T/2)\Psi_0^t(\tau)\right] \Psi_L(\tau).
\label{EQ:ratio}
\ee  
The integral in (\ref{EQ:ratio}) may be  evaluated  using
only the asymptotic behaviour of $\Psi_0$ and $\Xi_0$,
which involve $\zcl$ and $\bzcl$. This asymptotic behaviour
depends only on the form of the Hamiltonian in the
neighbourhood of the endpoints.

In all cases we  consider   the instanton solutions  have
the property  that $\bzcl = \zcl^* $ at their endpoints.
Here the asterisk denotes a true complex conjugate as
opposed to the formal conjugate  denoted by the bar. The
coincidence of the formal and true conjugate occurs 
because these  endpoints lie on the real unit 
sphere\footnote{This is a not a trivial observation: in  a more
accurate representation of the  Fe$_8$ problem which
includes  fourth order anisotropy terms there are
additional instantons for which this fails to be true.}.
Taking this observation into account, we   parametrize the
Hamiltonian in  the vicinity of the initial stationary
point in terms of two frequencies, $\omega_{1,2}$, as 
\be
H(\bz,z) \approx  {2j \over (1+z_i^* z_i)^2}
         \Bigl[ \omega_1(z-z_i)(\bz - z_i^*)
                +\tshf\omega_2 (z - z_i)^2
                +\tshf\omega_2^*(\bz - z_i^*)^2 \Bigr]. 
\label{EQ:Hend}
\ee
Since  $H(\bz,z)$ is real,  
so is $\omega_1$. 
Also, because  the initial point is an energy minimum, we must have   
$\omega_1 > |\omega_2|$. 
We can therefore define a real mean frequency,  $\omega$, by 
\be
\omega^2 \equiv \omega_1^2 - \omega_2\omega_2^*. 
\label{EQ:tom123}
\ee
A similar expression holds at $z_f$ with the same values of $\omega_1$
and $\omega_2$ provided the degeneracy in the Hamiltonian is due to some
symmetry. (There  might be   an extra phase factor in
$\omega_2$, but this makes no difference to the subsequent
calculation).

As $\tau$ becomes
large and negative, $B \to  \omega_2$,  $\bar B \to \omega^*_2$ and
$A=\phi_{SK} \to \omega_1$,
so we  see that 
\be
\left(\matrix{ \psi_0 \cr\bpsi_0}\right) \to 
\left(\matrix{ \psi_{0-} \cr\bpsi_{0-}}\right) e^{\omega
\tau} ;\quad 
\left(\matrix{ \xi_0 \cr\bxi_0}\right) \to 
\left(\matrix{ \xi_{0-} \cr\bxi_{0-}}\right) e^{-\omega
\tau},
\ee
where
\be
\left[\matrix{\omega_2 & -\omega +\omega_1 \cr
              \omega+\omega_1 & \omega_2^*}\right]  
\left(\matrix{ \psi_{0-} \cr\bpsi_{0-}}\right)=0.
\ee There is an analogous relation for $(\xi_{0-}, \bxi_{0-})^t$.
We can use the Wronskian to connect $\Psi_{0-}$ with
$\Xi_{0-}$, so everything can be expressed in terms of $W$ and
the normalization $g$. Similar remarks apply to $\Psi_{0+}$ and
$\Xi_{0+}$. If we write
\be
 \left(\matrix{ \psi_{L} \cr\bpsi_{L}}\right)=
\alpha \left(\matrix{ \psi_0 \cr\bpsi_0}\right)
+\beta \left(\matrix{ \xi_0 \cr\bxi_0}\right),
\ee
and apply the boundary condition at $-T/2$ we can find $\alpha$ and
$\beta$, and hence 
\be
\left(\matrix{ \psi_{L}(\tau) \cr\bpsi_{L}(\tau)}\right)
= \frac 1W \left[ - \xi_{0-} e^{\omega T/2}\left
(\matrix{ \psi_0(\tau) \cr\bpsi_0(\tau)}\right)+
\psi_{0-} e^{-\omega T/2}\left(\matrix{ \xi_0(\tau)
\cr\bxi_0(\tau)}\right)\right].
\ee
Inserting this into (\ref{EQ:ratio}) and noting that the
$\psi_0$ $\bpsi_0$ terms dominate, we find
\be
\frac{\bpsi_L(T/2)}{\lambda_0(T)} = - \frac 1{W^2}
\xi_{0-}\bxi_{0+} e^{\omega T}\int_{-T/2}^{T/2} 
\left(\psi_0^2 +\bpsi_0^2\right)d\tau,    
\ee
or, 
\be
\frac{\bpsi_L(T/2)}{\lambda_0(T)}= 
\frac {|\omega_2|^2}{\psi_{0-}\bpsi_{0+}} \frac{e^{\omega
T}}{4\omega^2}.
\ee

Thus the one-instanton contribution to the propagator is
\be
K(\z_f, z_i, T) = \exp\left\{S_{\rm cl} + \tshf
\int_{-T/2}^{T/2} \phi_{\rm
SK} d\tau\right\} \sqrt{\frac{j}{\pi g}} \left[\frac
{\psi_{0-}\bpsi_{0+}}{|\omega_2|^2}\right]^{\tshf} 
(2\omega T e^{-\tshf \omega T}).
\ee
Note that $\psi_0$, $\bpsi_0$ are proportional to $\sqrt g$, thus
$\sqrt{g}$ drops out and we can simply put $g=1$ in the
sequel. 

Let 
\bea
\dot \zcl &\approx& \omega \zeta_- e^{\omega \tau}, \quad
\tau \to -\infty\nonumber\\
 \dot \bzcl &\approx& \omega \bzeta_+ e^{-\omega \tau}, \quad
\tau \to +\infty, 
\eea 
then
\be
 \psi_{0-}\bpsi_{0+}= \frac{\omega^2 \zeta_-\bzeta_+}{N}
\ee
 with 
 \be
 N= (1+\z_i z_i)(1+ \z_f z_f).
 \ee
 Using this we can write
 \be
 K(\z_f, z_i, T) = \exp\left\{S_{\rm cl} + \tshf
 \int_{-T/2}^{T/2} \phi_{\rm
SK} d\tau\right\} \sqrt{\frac{j}{\pi N}} \left[\frac
{\bzeta_+\zeta_-}{|\omega_2|^2}\right]^{\tshf} 
(2\omega^2 T e^{-\tshf \omega T}).
\ee

\section{Extracting the Energy Splitting}

Again assume that the coherent states  $\ket{z_i}$ and
$\ket{z_f}$ represent  spins pointing  along the directions
of  two equal energy  global minima of the Hamiltonian
$\hH$. Let $\ket{\psi_{i,f}}$ be the approximate
(tunnelling-ignored)  energy eigenstates  localized near
these minima. These should have their  phases chosen  so that when
tunnelling {\it is\/} included the    eigenstates become the linear
combinations 
\be
\ket{\psi_{\pm}} = {1 \over \root2}
                   (\ket{\psi_i} \pm \ket{\psi_f}).
\ee
If the 
energies of these states are
\be
E_{\pm} = E_{\rm av} \pm \tshf\Delta,
\ee
and 
$
a_{\alpha} \equiv \brak{z_{\alpha}}{\psi_{\alpha}},
$
then as $T$ becomes large  the coherent-state propagator
\be
K(\bz_f, z_i, T) = \eval{z_f}{e^{-\hH T}}{z_i},
\ee
is given by
\bea
K(\bz_f, z_i, T) &\approx& a_f a_i^* e^{-E_{\rm av}T}
                         \sinh(\tshf \Delta T),\nonumber\\
&=& a_f a_i^* e^{-E_{\rm av}T} 
\left(\frac 12 \Delta T + \frac 16 \frac{\Delta^3 T^3}{2^3}
+\cdots\right).
\eea
We will  find the energy splitting,  $\Delta$, by evaluating
$K$ in the one-instanton approximation and comparing with this expression.

It is necessary to find expressions for the amplitudes $a_i$
and $a_f$. These are obtained by looking at 
\be
K_f=\eval{z_f} {e^{-\hat H T}}{z_f} \approx |a_f|^2 e^{-E_{\rm av}
T},
\ee
and 
\be
K_i= \eval{z_i} {e^{-\hat H T}}{z_i} \approx |a_i|^2 e^{-E_{\rm av}
T},
\label{EQ:harmonic_approx}
\ee
both evaluated in the harmonic approximation. This
evaluation  is performed in the appendix. This results in   
\be
K_{f}= (1+ \z_f z_f)^{2j}
\sqrt{\frac{2\omega}{\omega+\omega_1}}
e^{-\tshf(\omega-\omega_1)T}  
\ee
and a similar expression for $K_{i}$.
Thus
\be
\tshf \Delta 
= \frac{e^{S_{\rm cl} + \tshf \int_{-T/2}^{T/2} (\phi_{\rm
SK}-\omega_1) d\tau}}{[(1+\z_f z_f)^j(1+\z_i z_i)^j]}  
\sqrt{\frac{j}{\pi N}} [2\omega(\omega+\omega_1)]^{\tshf}
\omega \left[\frac
{\bzeta_+\zeta_-}{|\omega_2|^2}\right]^{\tshf}.
\ee

Now 
\be
\frac {2\omega(\omega+\omega_1)}{\omega_2^2} =
\frac {2\omega}{\omega_1-\omega}
\ee
so finally 
\be
\Delta = 2\omega \sqrt P e^I,
\label{EQ:Dtagen}
\ee
where
\be
P= \frac{j}{\pi N}\frac {2\omega}{\omega_1-\omega}\bzeta_+\zeta_-
\ee
and
\be
I= j \int_{-\infty}^{\infty} 
a_{\rm wz}(\tau) \,d\tau + \tshf \int_{-\infty}^{\infty}(\phi_{\rm
SK}-\omega_1) d\tau,
\ee
where $a_{\rm wz}$ is the kinetic term 
\be
a_{\rm WZ}(\tau) = {{\dot\bzcl} \zcl - {\dot\zcl}\bzcl
                     \over 1 + \bzcl\zcl}
\ee
in the classical
action --- the boundary terms having cancelled with the 
$(1+\z_f z_f)^j(1+\z_i z_i)^j$ in the denominator.

\section{The LMG model.}

In this section we will evaluate the tunnel splitting in the
relatively simple case of the Lipkin-Meshkov-Glick (LMG)
model\cite{lipkin65}.

We will take the  LMG Hamiltonian to be
\be
\hH = {w \over \root2 (2j -1)}(\hat J_+^2 + \hat J_-^2) + {jw \over \root2},
\ee
with $w > 0$. Since $\hat J_+^2 + \hat J_-^2 = 2(\hat J_x^2 -
\hat J_y^2)$, we see that
the classical minima lie along $\pm\yhat$. 
The Hamiltonian which appears in the path integral is    
\be
H(\bz,z) = \frac{\eval{z}{\hat H}{z}}{\brak{z}{z}}=\root2 j w  {z^2 + \bz^2 \over
                        (1 + \bz z)^2} + {jw \over \root2}. \label{EQ:hamlmg}
\ee
By setting $\partial H/\partial z = \partial H/\partial\bz =0$, 
the classical minima are
found to be at the points
\be
(z,\bz) = (i,-i),\quad (-i,i),
\ee
which correspond to the $\pm\yhat$ directions of the Cartesian
axes. The explicitly added constant in $\hH$ is chosen 
to make $H(\bz,z)$  zero at these points.

Now we write down the equations of motion for the instantons 
\bea
{\dot\bz} &=& \root2 w {z - \bz^3 \over (1 + \bz z)}, \nnu\\
{\dot z} &=& -\root2 w {\bz - z^3 \over (1 + \bz z)}.
\label{EQ:LMGinsteq}
\eea
We seek a solution  which goes from $(z_i,\bz_i) = (-i,i)$ to
$(z_f,\bz_f) = (i,-i)$.  The two equations in (\ref{EQ:LMGinsteq}) 
can be decoupled  by
exploiting the energy conservation condition  $H(\bz,z) =
0$ which follows from the Hamiltonian nature of the
trajectory. This can be written as
\be
2(z^2 + \bz^2) + 1 + 2\bz z + \bz^2 z^2 = 0.
\ee
and may be solved to yield $z$ as a  function of $\z$ and {\it
vice
versa\/}:
\be
\bz = -i {\root2 z + i \over z + \root2 i}, \quad
z = -i {\root2 \bz + i \over \bz + \root2 i}.
\ee
(Choosing the  
other solution of the quadratic equation yields instantons running in the opposite direction.)
Substituting these formulae  in the equations of motion yields
\be
\dot\bz = -i w (1+\bz^2), \quad \dot z = i w (1 + z^2).
\label{EQ:eom_lmg_2}
\ee

These may be integrated by elementary means to yield
\bea
\zcl(\tau) &=& i {e^{2w\tau} - C \over e^{2w\tau} + C}
                 = i \tanh w(\tau - \tau_0), \label{EQ:lmg_zcl}\\
\bzcl(\tau) &=& -i {e^{2w\tau} - C' \over e^{2w\tau} + C'}
                 = -i \tanh w(\tau - \tau'_0),  \label{EQ:lmg_bzcl} 
\eea
where $C = e^{2w\tau_0}$, $C' = e^{2w\tau'_0}$. These
constants are not independent.  Energy conservation
requires
\be
{C' \over C} = {\root2 - 1 \over \root2 + 1}.
\ee

It is useful at this point to find the
frequencies $\omega$, $\omega_1$ and $\omega_2$. We have
\be
\omega_1 = {(1 + \bz_i z_i)^2 \over 2j}
         \left.{\partial^2 H \over \partial z \partial\bz}\right|_i,
                   \quad
\omega_2 = {(1 + \bz_i z_i)^2 \over 2j}
         \left.{\partial^2 H \over \partial z^2}\right|_i,
\ee
where the suffix $i$ means that the derivatives are to be evaluated at the
initial point. Carrying out the algebra, we obtain
\be
\omega_1 = {3 \over \root2} w, \quad \omega_2 = {1\over \root2} w.
\ee
Hence,
\be
\omega = (\omega_1^2 - \omega_2^2)^{1/2} = 2w.
\ee

We can now evaluate
the Wess-Zumino and Solari-Kochetov terms in the tunnelling action. We denote
these by $I_{\rm WZ}$ and $I_{\rm SK}$ respectively.  
We begin with $I_{\rm WZ}$. 
If we make use of \eno{EQ:eom_lmg_2},
we find
\be
a_{\rm WZ}(\tau) = {1\over 1 +\bz z} (\dot\bz z - \bz \dot z)
                 = -i w (\bz + z).
\ee
Substituting the explicit forms and performing the integration we get
\be
I_{\rm WZ} = -j \ln(C/C') = -2j \ln(1+\root2).
\ee

Now we integrate
the Solari-Kochetov term along the instanton trajectory. 
We find that 
\be
\phi_{\rm SK} = -\root2 {(z^2 + \bz^2)(2 + \bz z) \over
                         (1 + \bz z)^2}.
\ee
By energy conservation  this equals
\be
{w \over \root2}(2 + \bz z).
\ee
Thus, along the instanton,
\be
\phi_{\rm SK} - \omega_1 = -{w \over \root2}(1 - \bz z) 
                      = -iw (\bz + z).
\ee
Hence
\be
I_{\rm SK} = \hf \int_{-\infty}^{\infty} (\phi_{\rm SK} - \omega_1) d\tau
           = - \ln (1+\root2).
\ee
The total tunnelling action is
\be
I = -(2j + 1) \ln(1 + \root2),
\ee
The shift $2j \to 2j+1$ is due to the Solari-Kochetov
correction. 

We must now evaluate  $P$.
This consists of a product of various factors, all of which
are to hand. Thus,
\be
{j \over \pi N} = {j \over 4\pi}.
\ee
The factors $\bzeta_+$ and $\zeta_-$ are found by differentiating the
formulas (\ref{EQ:lmg_zcl}) and (\ref{EQ:lmg_bzcl}) and examining the limits
$\tau\to\pm\infty$. In this we way we get
\be
\zeta_- \bzeta_+ = 4{C' \over C}.
\ee
Finally,
\be
{2\omega \over \omega_1 - \omega} = 4\root2 (3 + 2\root2).
\ee
Putting these together, we obtain
\be
P = -{4 j \over \pi} (4 + 3\root2) {C' \over C} = {4j \over \pi} \root2.
\ee

At this point we have almost all that we need to write down the answer for
the tunnel splitting --- except that we need to consider a
second instanton.
The trajectory (\ref{EQ:lmg_zcl}) and (\ref{EQ:lmg_bzcl}) passes
close to the
north pole of the sphere. By symmetry there must be a second instanton 
which passes near
the south pole. This  is given by
\be
\zcl = i \coth w(\tau - \tau_0),\quad
\bzcl = -i \coth w(\tau - \tau'_0).
\ee
It is obvious by symmetry again that this
instanton has exactly the same amplitude as the first, so the total
amplitude (and thus the splitting) is obtained by simply doubling the
answer from the first instanton. Hence 
\be
\Delta = 16 w \lf( j \over \pi\rt)^{1/2} 2^{1/4}
        e^{-(2j + 1) \ln(1+\root2)}.
\ee
This agrees with \cite{belinicher97,enz86,garg98}
[In the last reference put $\xi^2 = 1/\root2$ in Eqs. 4.31--4.34.]
In \cite{garg98} there  a numerical comparison which shows that
the prefactor is indeed correct.

For completeness, we note that the  average energy   is 
given by $E_{\rm av}= \tshf(\omega - \omega_1)$.

\section{Application to Fe$_8$}

The LMG model is of interest primarily because it provides
a check of our formalism  against other well-confirmed
calculations. In  this section we will calculate the tunnel
splitting for  a family of models   that includes  a
realistic  approximation to  the molecular magnet \Fe8.   
The spin-direction dependent energy  in  \Fe8 is less symmetric
than that of the LMG, and the relevant Hamiltonian 
includes  an externally imposed magnetic field which serves
to pull the classical minima off the equator of the unit
sphere. It is the experimentally observed oscillations in
the tunnel splitting as a function of the external field that makes
this system interesting. The oscillations are a 
consequence of interference between the two distinct
instanton trajectories and are accurately reproduced by our
calculation.

We take as our Hamiltonian
\be
\hH = k_1 \hat J_z^2 + k_2 \hat J_y^2 - g \mu_B H \hat J_z,
\ee
with $k_1 > k_2 > 0$. We define $\lam = k_2/k_1$, $H_c = 2k_1 j/g\mu_B$
and
\be
h = H/H_c.
\ee
We will express all results in terms of the combinations $\lam$ and $h$. 
It is also convenient  to define a  $1/j$ corrected field $\htil$, and anisotropy $\ktil$
by
\be
\htil = j h /(j - \tshf), \quad \ktil = k (j - \tshf)/j. 
\ee

We follow the same steps as in the LMG model. 
The   ``classical'' Hamiltonian appearing in the path integral is
\be
H(\bz,z) = \frac{\eval{z}{\hat H}{z}}{\brak{z}{z}}=
\ktil j^2
            \lf[(1-\bz z)^2 - \lam (z - \bz)^2 - 2\htil (1 - \bz^2 z^2)
                   \over
                 (1 + \bz z)^2 \rt]. 
\label{EQ:hamfe8}
\ee
The energy minima are
now  at the points
\be
\bz = z = \pm z_0,
\ee
where $z_0$ is real and given by
\be
z_0 = [(1-\htil)/(1+\htil)]^{1/2}.
\ee
In Cartesian coordinates these minima lie in the {\it
xz\/} plane --- provided  we confine ourselves to $\htil < 1$, which we shall do.
In fact, we will assume that
\be
\htil < \rtlb. 
\label{EQ:hupper}
\ee
At the two minima, the energy is
\be
\eps_0 = H(\bz_0, z_0) = -\ktil j^2 \htil^2.
\ee

The classical equations of motion 
are 
\bea
{\dot\bz} &=& {\ktil j \over (1 + \bz z)}
              \lf[ -2 \bz (1 - \bz z) + \lam (\bz - z) (1 + \bz^2)
                              + 2 \htil \bz (1 + \bz z)\rt], \nnu\\
{\dot z} &=& -{\ktil j \over (1 + \bz z)}
              \lf[ -2 z (1 - \bz z) + \lam (z - \bz) (1 + z^2)
                              + 2 \htil z (1 + \bz z)\rt].
\eea
We wish to solve these subject to the boundary conditions
$z_i = z(-\infty) = z_0$, $\bz_f = z(\infty) = -z_0$. Note that
$\bz_i = z_i$, $z_f = \bz_f$, so the instanton end points
still lie
on the real sphere, but the rest of the instanton does not.
Once again the
equations can be decoupled by exploiting the fact that energy is conserved
along the instanton trajectory. In this case $H(\bz, z) = \eps_0$. 
This condition can be
written as
\be
(1 - \bz z)^2 - \lam (z - \bz)^2 - 2 \htil (1 - \bz^2 z^2)
          = -\htil^2 (1 + \bz z)^2,
\ee
and may be    solved to give
\be
\bz = {\rtl z \pm (1 - \htil) \over \rtl \pm (1 + \htil) z}. 
\label{EQ:zbar(z)}
\ee
Substituting this in the equation of motion for $\dot z$, and
simplifying, we get
\be
\dot z = \pm \rtl (1 + \htil) \ktil j (z_0^2 - z^2). 
\label{EQ:dotz}
\ee
We will see that to obtain instantons going from $z_0$ to $-z_0$,
we must pick the minus sign in this equation. The other sign yields
instantons running in the opposite direction.

It is now elementary to integrate \eno{EQ:dotz}, and use \eno{EQ:zbar(z)} to
obtain the time dependence for both $\zcl(\tau)$ and
$\bzcl(\tau)$. We find 
\bea
\zcl(\tau) &=& -z_0 \tanh t,  \label{EQ:zcl} \\
\bzcl(\tau) &=& -z_0 { \rtl\tanh t + \rtb 
                       \over \rtl + \rtb\tanh t}. 
\label{EQ:bzcl}
\eea
Here,
\be
t = \omega\tau/2,
\ee
and the frequency $\omega$ is given by
\be
\omega   = 2\ktil j [\lam (1 - \htil^2)]^{1/2}. \label{EQ:tom}
\ee
That this is the same $\omega$ that follows from
Eqs.~(\ref{EQ:Hend})
and (\ref{EQ:tom123}) shall be shown shortly. It can be seen that our
solution corresponds to choosing the minus sign in \eno{EQ:dotz} as
asserted above. It is also useful to note that the solution
(\ref{EQ:zcl})
and (\ref{EQ:bzcl}) can be rewritten as
\be
\zcl = -z_0 \tanh t, \quad \bzcl = -z_0 \coth(t+t_0),
\ee
where
\be
\tanh t_0 = \lf( {\lam \over 1 - \htil^2} \rt)^{1/2}.
\ee

Equations (\ref{EQ:dotz}) and (\ref{EQ:zbar(z)}) possess a second solution,
\be
\zcl = -z_0 \coth t, \quad \bzcl = -z_0 \tanh(t+t_0).
\ee
 Formally, this new trajectory  can be
obtained from the first by the shift $t \to t + i\pi/2$. Alternatively,
we could obtain it by switching the expressions for
$\zcl$ and $\bzcl$ in Eqs. (\ref{EQ:zcl}) and (\ref{EQ:bzcl}), which corresponds
to reflection in the {\it xz\/} plane --- a symmetry of
the Hamiltonian --- and then shifting $t$ by $-t_0$.

Again we  find the
frequencies $\omega$, $\omega_1$ and $\omega_2$. We note that
\be
\omega_1 = {(1 + \bz_i z_i)^2 \over 2j}
         \lf.{\partial^2 H \over \partial z \partial\bz}\rt|_i,
                   \quad
\omega_2 = {(1 + \bz_i z_i)^2 \over 2j}
         \lf.{\partial^2 H \over \partial z^2}\rt|_i,
\ee
where the suffix $i$ means that the derivatives are to be evaluated at the
initial point $\bz = z = z_i$. Carrying out the algebra, we obtain
\bea
\omega_1 &=& \ktil j (1 - \htil^2 + \lam), \\ 
\omega_2 &=& \ktil j (1 - \htil^2 - \lam).
\eea
We now  use \eno{EQ:tom123} to show that $\omega$ is
given by \eno{EQ:tom}. The same frequencies are found at the final point
$\bz = z = z_f$.

We  next evaluate and integrate the
Wess-Zumino and Solari-Kochetov terms in the tunnelling action. We denote these
by $I_{\rm WZ}$ and $I_{\rm SK}$. Since the calculations are somewhat lengthy,
it is best to do the two terms separately. We begin with $I_{\rm WZ}$, considering
instanton 1, i.e., that given by (\ref{EQ:zcl}) and
(\ref{EQ:bzcl}). 
After some algebra, we obtain
\be
a_{\rm WZ}(\tau) = -{\pi_2(\tanh t) \over \pi_3 (\tanh t)} {\omega \over 2}
                     \sech^2 t,
\ee
where $\pi_2$ and $\pi_3$ are polynomials of degree 2 and 3, whose explicit
form we do not require. What we do need is the differential $a_{\rm WZ}\,d\tau$.
If we make the substitution
\be
v = \tanh t,
\label{EQ:ttov}
\ee
and factorize the polynomials $\pi_2$ and $\pi_3$, we obtain
\be
\int_{-\infty}^{\infty} a_{\rm WZ}(\tau) d\tau
   = - \int_{-1}^ 1 { (v - v_3)(v - v_4) \over (v-v_1)(v-v_2)(v-v_5)} dv,
     \label{EQ:intwz_v}
\ee
where
\bea
v_{1,2} &=& {1 \over \rtl} \lf( {1 + \htil \over 1 - \htil } \rt)^{1/2}
              \lf( -1 \pm \rtlb \rt),  \\
v_{3,4} &=& { - \rtb \pm \sqrt{1 - \htil^2 - \lam} \over \rtl},  \\
v_5     &=& -{\rtl \over \rtb}.
\eea
The integral is best done by decomposing the integrand into partial
fractions. We find
\be
{(v - v_3)(v - v_4) \over (v-v_1)(v-v_2)(v-v_5)} = 
 {1 \over v - v_5} + {\beta \over v - v_1} - {\beta \over v - v_2},
\label{EQ:parfrac}
\ee
where
\be
\beta = - {\htil \over \rtlb}.
\ee
Thus,
\be
I_{\rm WZ} = j \int_{-\infty}^{\infty} a_{\rm WZ} d\tau
           = -j \lf[ \ln\rat5 + \beta\ln\rat1 - \beta\ln\rat2 \rt].
\ee
The ratio involving $v_5$ is
\be
{1 - v_5 \over -1 - v_5} = {\rtl + \rtb \over \rtl - \rtb} 
                        \equiv \Rtil_1,
\ee
while the $\beta$ terms combine to yield the logarithm of
\be
{1 - v_1v_2 + (v_2 - v_1) \over 1 - v_1v_2 - (v_2 - v_1)}
    = {\htil\rtl + \sqrt{1-\lam}\rtb
                    \over \htil\rtl - \sqrt{1-\lam}\rtb }
    \equiv \Rtil_2.
\ee
Collecting together the various parts, we have
\be
I_{{\rm WZ},1} = -j  \ln\Rtil_1 +
           {j \htil \over \rtlb} \ln\Rtil_2.
\ee
We have added another suffix to show that this pertains to
instanton 1.

The next step is to integrate
the Solari-Kochetov term. For this we first need $\phi_{\rm SK}$.
From Eqns.~(\ref{EQ:hamfe8}) and (\ref{EQ:defsk}) we find,
\bea
\phi_{\rm SK} = {\ktil j \over 2 (1 +\bz z)^2}
             \Bigl[&&-4\lf(1 - 2\bz z - (\bz z)^2 \rt)
                + 2\lam\lf( (1+\bz z)^2 + 2(\bz -z)^2 + \bz z(\bz -z)^2\rt)
                   \Bigr. \nnu\\
                &&+ 4 \Bigl.\htil (1 + \bz z)^2 \Bigr].
\eea
(The reader may verify that as $\tau \to \pm\infty$,
$\phi_{\rm SK} \to \omega_1$. This provides a check on our earlier
calculation of $\omega_1$.) After a little more work, we find,
\bea
\phi_{\rm SK} - \omega_1 &=& {\ktil j \over (1 +\bz z)^2}
                 \biggl[ (1+\htil)^2 (1+ \bz z)^2 +
                           \lam (\bz-z)^2 -4 \biggr] \nnu \\
               &&\ + \ktil j\lam {(\bz - z)^2 \over 1 + \bz z}.
\eea
We have written this expression in such a way that it is convenient to
integrate the terms on the two lines separately. That is, we write,
\be
I_{\rm SK} = \hf \int_{-\infty}^{\infty} (\phi_{\rm SK} - \omega_1) d\tau
           = I_A + I_B,
\ee
where,
\bea
I_A &=& \hf \ktil j \int_{-\infty}^{\infty} d\tau
                 {(1+\htil)^2 (1+ \bz z)^2 + \lam (\bz-z)^2 -4 
                   \over (1 + \bz z)^2}, \\
I_B &=& \hf \ktil j \lam \int_{-\infty}^{\infty} d\tau 
         {(\bz - z)^2 \over 1 + \bz z}.
\eea
It is to be understood that the integrands are evaluated
along the instanton trajectories.
We may exploit this fact to simplify the integrand for $I_A$ by using
energy conservation and so  eliminate the term in $\lam$. When this is done,
we obtain
\be
I_A = \ktil j (1+\htil) \int_{-\infty}^{\infty} d\tau
         {-(1-\htil) + (1+\htil)\bz z \over 1 + \bz z}.
\ee

The integrals are evaluated in the same way as $I_{\rm WZ}$.
With the same change of variables, and definitions of $v_1$ to $v_5$ as
before, for $I_A$ we get
\bea
I_A &=& - {(1-\htil^2)^{1/2} (1 +\htil) \over \rtl (1 - \htil)}
            \int_{-1}^1 {dv \over (v -v_1) (v - v_2)} \nnu \\
    &=& - {(1 + \htil) \over 2 \rtlb}
            \int_{-1}^1 \lf[ {1\over v-v_1} - {1\over v-v_2}\rt]\,dv \nnu\\
    &=& - {(1 + \htil) \over 2 \rtlb} \ln\Rtil_2.
\eea
Likewise, for $I_B$ we get
\be
I_B = -\hf \int_{-1}^1
          {(v-1)(v+1) \over (v-v_1)(v-v_2)(v-v_5)}\, dv
\ee
The partial fraction decomposition yields
\be
{(v-1)(v+1) \over (v-v_1)(v-v_2)(v-v_5)} = 
   {1 \over v - v_5} + {\beta' \over v - v_1} - {\beta' \over v - v_2},
\ee
where
\be
\beta' = - (1 - \lam)^{-1/2}.
\ee
Hence,
\be
I_B = -\hf \ln\Rtil_1 + {1 \over 2 \rtlb} \ln\Rtil_2.
\ee

Thus the Solari-Kochetov contribution to the action for instanton 1 is
\be
I_{\rm SK} = -\hf \ln\Rtil_1 - {\htil \over 2 \rtlb} \ln\Rtil_2.
\ee
Note that this is $O(1/j)$ relative to the Wess-Zumino contribution.
Adding together the two contributions, we obtain the total action
\be
I = -{(j + \tshf)} \ln\Rtil_1 + {j h \over \rtlb}\ln\Rtil_2.
\label{EQ:Iinst1}
\ee
In the second term we have used the formula $(j - \hf)\htil = j h$.

We now turn to the prefactor  $P$.
In evaluating this, we may ignore differences of order $1/j$, i.e., we may
replace $\jtil$ by $j$, $\htil$ by $h$, etc. The quantity consists of
a product of various factors, all of which are already available. Thus,
\be
{j \over \pi N} = {j \over \pi (1+z_0^2)^2}.
\ee
The factors $\bzeta_+$ and $\zeta_-$ are  found by differentiating the
formulas (\ref{EQ:zcl}) and (\ref{EQ:bzcl}) and examining the limits
$\tau\to\pm\infty$. In this we way we get
\bea
\zeta_- &=& -2z_0, \\
\bzeta_+ &=& 2z_0 {\rthb - \rtl \over \rthb + \rtl}.
\eea
Finally,
\be
{2\omega \over \omega_1 - \omega} = 
                   4 {\sqrt{\lam (1 - h^2)} \over (1 - h^2 + \lam)
                                - 2\sqrt{\lam (1-h^2)}}
      = 4 {\sqrt{\lam(1 - h^2)} \over \lf[\rthb - \rtl\rt]^2}.
\ee
Making use of the identity
\be
{2z_0 \over 1 + z_0^2} = (1 - h^2)^{1/2},
\ee
we obtain
\be
P = -{4 j \over \pi} {\lam^{1/2} (1-h^2)^{3/2}
                       \over 1 - h^2 - \lam}. \label{EQ:Pinst1}
\ee

We can now obtain the
contribution of instanton 1 to the tunnelling amplitude by substituting
Eqs. (\ref{EQ:Iinst1}) and (\ref{EQ:Pinst1}) in the general formula
(\ref{EQ:Dtagen}). Denoting this quantity by $\Delta_1$, we have
\be
\Delta_1 = 2 \omega \sqrt{|P|} e^{I - i\pi/2},
\ee
where the additional factor of $e^{-i\pi/2}$ arises from the fact that
$P < 0$.

It remains to obtain the
tunnelling amplitude $\Delta_2$ from the second instanton. Because the two
instantons are related by a complex shift in $t$, it is apparent that
the actions $I_{1,2}$ (where we temporarily add suffixes to distinguish
the two) and the prefactors $P_{1,2}$ will be given
by the same analytic expressions. However, the phases to be assigned to
the actions and $\sqrt{P}$ are somewhat ambiguous. Unlike the case of a
particle moving in one dimension, the prefactor in the general formula
does not arise as the determinant of a Hermitean quadratic form, and
there is no unambiguous way for factors of $i$ to get partitioned between
the prefactor and the exponent. The surest way of fixing the relative
phases is to appeal to a physical argument. Alternatively, this can be
regarded as fixing the signs of the amplitudes $a_i$ and $a_f$.

For the \Fe8 Hamiltonian (\ref{EQ:hamfe8}), let us work in the
$J_z$ basis $\ket{j,m}$ with the standard definition of the raising
and lowering operators $J_{\pm}$, so that the matrix elements
$\mel{j,m\pm 1}{J_{\pm}}{j,m}$ are all real. Then the matrix
of $\hH$ is
completely real, and since it is Hermitean, all its eigenvalues and
eigenvectors are also real. Secondly, since $z_i = z_0$ and $z_f = -z_0$ are
real, the states $\ket{z_{i,f}}$ are real, i.e., all the matrix elements
$\brak{j,m}{z_{i,f}}$ are real. Thus the amplitudes $a_i$ and $a_f$ are
real. It follows that the amplitude $K$ is real, and so is the
one-instanton contribution to it, i.e., $\Delta_1 + \Delta_2$ is real.
Therefore, we must have
\be
\Delta_2 = \Delta_1^*. \label{EQ:D12}
\ee

Equation (\ref{EQ:D12}) determines $\Delta_2$, and the energy splitting $\Delta$
completely. However, it is still useful to investigate the origin of
the phase difference in the actions a little more closely. As readers
will have noticed already, the integrand in \eno{EQ:intwz_v} is
singular at $v = v_2$ and $v = v_5$, since for $\htil < \rtlb$,
\be
v_1 < - 1, \quad -1 < v_2 < 1, \quad -1 < v_5 < 1.
\ee
Correspondingly, both $\Rtil_1$ and $\Rtil_2$ are negative, and both
$\ln\Rtil_1$ and $\ln\Rtil_2$ must be interpreted to have an imaginary
part of $\pi$ modulo an integer multiple of $2\pi$. The question is
what the assignment should be for the two instantons. We can see this
most easily by examining the difference
$\Delta I_{\rm WZ} = I_{{\rm WZ},2} - I_{{\rm WZ},1}$.
To this end, we note that the WZ one-form may be
written as a complex one-form in the $z$ plane,
\be
a_{\rm WZ}\, d\tau = {1 \over 1 + z \bz(z)}
                  \lf[ z {d\bz \over dz} - \bz(z) \rt] dz
                   \equiv F(z) dz,
\ee
with $\bz(z)$ given by \eno{EQ:zbar(z)}. Thus, $I_{\rm WZ}$ may
be written as a $z$-plane contour integral of $F(z)$ from $z_0$ to
$-z_0$. In fact, apart
from a scale factor of $z_0$, the substitution (\ref{EQ:ttov}) is 
tantamount to changing the integration variable to $z$, so we
see that $F(z)$ has poles at $z_0 v_2$ and $z_0 v_5$ (the one at
$z_0 v_1$ does not matter). The two instantons go around these poles
in opposite senses, so $\Delta I_{\rm WZ}$ is given by integrating $F(z)$
along a closed contour from $z_0$ to $-z_0$ and back to $z_0$:
\be
\Delta I_{\rm WZ} = \oint F(z) dz.
\ee
The residues at the poles can be read off the partial fraction
decomposition (\ref{EQ:parfrac}), yielding
\be
I_{{\rm WZ},2} - I_{{\rm WZ},1}
           = 2j\pi - {2j\htil\pi \over \rtlb}.
\ee
This is precisely what we would obtain from \eno{EQ:D12}, for that would
have us assign $\pm i\pi$ for $\ln\Rtil_1$ (and $\ln\Rtil_2$) for the
two instantons.

The energy splitting is given
by
\be
\Delta = \Delta_1 + \Delta_2^*.
\ee
To compare with previous results, it is useful to rewrite this as
follows. Consider the real part of the action,
\be
\Gamma_0 = -{\rm Re}\,I
         =  (j + \hf) \ln|\Rtil_1|
              - {jh \over \sqrt{1-\lam}} \ln|\Rtil_2|.
\ee
The ratios $\Rtil_1$ and $\Rtil_2$ are defined in terms of the
field $\htil$. If we write $\htil = h + O(1/j)$, and expand in powers
of $1/j$, we discover that
\be
\Gamma_0 =  (j + \hf) \ln|R_1|
              - {jh \over \sqrt{1-\lam}} \ln|R_2| + O(j^{-1}),
\ee
where $R_i$ is obtained from $\Rtil_i$ by simply deleting the tildes
above the $h$'s. Note that the corrections are of $O(1/j)$, not $O(1)$.
These are beyond the accuracy to which we are working, so we simply drop
them henceforth.

Thus, the complete expression for the splitting is
\be
\Delta = {\sqrt{8\over \pi}}\omega F^{1/2} e^{-\Gamma_0}\cos\Lam.
\ee
We give the expressions for $F$, $\Gamma_0$ and $\Lam$ for ready reference:
\bea
F &=& 8 j {\lam^{1/2} (1-h^2)^{3/2} 
                       \over 1 - h^2 - \lam}, \\
\Gamma_0 &=& (j + \hf) \ln \lf[{\rthb + \rtl \over \rthb - \rtl}\rt]
           -{jh\over\rtlb}\ln
              \lf[{\sqrt{(1-\lam)(1 - h^2)} +  h\rtl
                    \over \sqrt{(1-\lam)(1 - h^2)} -  h\rtl}\rt], \\
\Lam &=& {\rm Im}\,I - {\pi \over 2} \nnu\\
       &=& j\pi \left( 1 - {h \over \sqrt{1-\lam}} \right).
\eea

Our answer for $\Delta$ is identical to that found by means
of the discrete WKB method in
\cite{garg01}  [see Eqs. (5.1--5.5)].
Naturally, the points at the which the tunnel splitting vanishes
are the same too.

The nontrivial aspect of this calculation is that there are $1/j$ corrections
in the quenching condition. If we simply take the energy expectation
$H(\bz, z) = \eval{z}{\hH}{z}/\brak{z}{z}$ in the Wess-Zumino term, we have the
problem that the anisotropy and field terms scale with $j$ differently if
$1/j$ corrections are included. This is how the quenching condition was
found in  \cite{garg93}, but the $1/j$ corrections were
never considered, so it was some what serendipitous that the condition that was
stated turned out to be rigorously correct. By including the SK correction,
this deficiency is now repaired.

\section{Conclusion}

We have seen that, once the extra-phase contribution is
included, the coherent-state path integral for spin
provides an accurate and effective tool for calculating
tunnel splitting. In particular the Weyl shift $j\to
j+\tshf$ appears automatically. It must therefore be
possible to put the spin coherent-state  path integral on
the same sound mathematical footing as the conventional
Feynman integral.   

\section{Acknowledgments}

Work at Urbana and Evanston was supported by the National Science
Foundation under grants DMR-98-17941 and DMR-9616749, respectively.
EK thanks the Russian Foundation for Fundamental
Research for supporting this work through Grant $N{\underline{o}}$ 00-01-00049.

\section{Appendix}

Here we derive equation (\ref{EQ:harmonic_approx}). We first
apply an $SU(2)$ rotation to   
\be
H_{\rm initial}(\bz,z) = {2j \over (1+z_i^* z_i)^2}
         \left[ \omega_1(z-z_i)(\bz - z_i^*)
                +\tshf\omega_2 (z - z_i)^2
                +\tshf\omega_2^*(\bz - z_i^*)^2 \right] 
\label{EQ:Hend2}
\ee
in order to  place  $z_i$, $\z_i$  at the origin, and
to make the coefficient $\omega_2$ real. The result is  
\be
H(\bz,z) = 2j 
         \left[ \omega_1\z z
                +\tshf\omega_2 z^2
                +\tshf\omega_2 \bz^2 \right].
\ee		
In the semiclassical limit, $2j\gg 1$, we may 
ignore the curvature of the phase space
and, after rescaling  $\sqrt{2j} \,z\to z$ to account for the
difference in the coefficient in the kinetic terms,  identify $H(\bz,z)$
with the coherent state classical Hamiltonian  for the squeezed
harmonic oscillator 
\be
\hat H = \omega_1\ad a +\tshf\omega_2({\ad}^2 +a^2).
\ee  
The Bogoliubov transformation
\bea
b&=& \cosh \theta \,a +\sinh \theta \,\ad\nonumber\\
\bd &=& \sinh \theta \, a +\cosh\theta \,\ad
\eea
reduces the Hamiltonian
\be 
\hat H_{\rm squeezed}= \Omega\cosh 2\theta \left(\ad a+\tshf\right) +\tshf
\Omega\sinh 2\theta ({\ad}^2+a^2)
\ee
to
\be
\hat H_{\rm squeezed}=\Omega\left(\bd b+\tshf\right),
\ee  
and so  we identify 
\bea
\Omega = \omega &=&\sqrt{\omega_1^2 -
\omega_2^2},\nonumber\\
 \Omega \cosh 2\theta&=& \omega_1,\nonumber\\
 \Omega \sinh 2\theta &=& \omega_2.
 \eea  
 The  eigenvalues of $\hat H$
are therefore 
\be
E_n = \omega(n+ \tshf) - \tshf \omega_1.
\ee
The operators $\ad a$, $a^2$ and
${\ad}^2$ generate the Lie algebra  $su(1,1)$.   Therefore either the 
flat phase-space coherent state path integral or     
standard $su(1,1)$  disentangling methods 
\cite{yuen76,stoler70} can be used to derive  
\be
\eval{\zeta_f}{e^{-\hat HT}}{\zeta_i} =
D^{-\tshf} \exp\left \{D^{-1}\left(\bzeta_f\zeta_i - \tshf \sinh
2\theta \sinh \omega
T(\bzeta_f^2+\zeta_i^2)\right)\right\}e^{-\tshf \omega_1T},
\ee
where
\be
D=e^{\omega T} \cosh^2\theta  - e^{-\omega T}\sinh^2 \theta,
\ee
and  the harmonic oscillator coherent states $\ket{\zeta}$
are  defined by
\be
\ket{\zeta}= \exp \zeta \ad\,  \ket{0}, \quad a \ket{0}=0.
\ee      
In the large$-T$ limit, and with $\zeta_i$ and $\bzeta_f$
both at the origin, this  reduces to 
\be       
\eval{0}{e^{-\hat HT}}{0} \to  (\cosh\theta)^{-1}
e^{-\tshf(\omega-\omega_1)T}= \sqrt{\frac{2\omega}{
\omega+\omega_1}}e^{-\tshf(\omega-\omega_1)T}.
\ee
We now rotate back to the  original $z_i$. Taking note of the
transformation properties of the $\ket{z}$'s, we get
\be 
K_i= (1+ \z_i z_i)^{2j}
\sqrt{\frac{2\omega}{\omega+\omega_1}}
e^{-\tshf(\omega-\omega_1)T},
\ee
as claimed.

%\begin{thebibliography}

\eject
\end{document}